\documentstyle[twocolumn,aps,graphicx]{revtex}
\def\be{\begin{equation}}
\def\ee{\end{equation}}
\def\eps{\epsilon}

\def\epseff{\epsilon}
\def\mueff{\mu}
\def\neff{n}

\begin{document}
\title{Effective medium theory of left-handed materials}

\author{T. Koschny and M. Kafesaki}
\address{Institute of Electronic Structure and Laser (IESL),
Foundation for Research and Technology Hellas (FORTH), \\
P.O. Box 1527, 71110 Heraklion, Crete, Greece.}

\author{E. N. Economou}
\address{IESL-FORTH, P.O. Box 1527,
71110 Heraklion, Crete, Greece, and Dept. Physics, University of
Crete, Greece}

\author{C.  M. Soukoulis$^*$}
\address{IESL-FORTH, and Dept. of Materials Science and
Technology, 71110 Heraklion, Crete, Greece}
\address{Ames Laboratory and Dept. Physics and Astronomy, Iowa State
University$^\dagger$, Ames, Iowa 50011}

\author{\parbox[t]{5.5in}{\small %
We analyze the transmission and reflection data obtained through
transfer matrix calculations on metamaterials of finite lengths,
to determine their effective permittivity $\eps$ and permeability $\mu$.
Our study concerns metamaterial structures composed of periodic
arrangements of wires, cut-wires, split ring resonators (SRRs),
closed-SRRs, and both wires and SRRs.
We find that the SRRs have a strong electric response,
equivalent to that of cut-wires, which dominates the behavior
of left-handed materials (LHM).
Analytical expressions for the effective parameters of the
different structures are given, which can be used to explain
the transmission characteristics of LHMs.
Of particular relevance is the criterion introduced by our studies
to identify if an experimental transmission peak is left- or right-handed.
 }}

\maketitle


Recently, there have been many studies about metamaterials that have a
negative refractive index $n$.
These materials, called left-handed materials (LHMs), theoretically
discussed by Veselago \cite{Veselago}, have simultaneously negative
electrical permittivity $\eps$ and magnetic permeability $\mu$.
Such materials consisting of split ring resonators (SRRs) and
continuous
wires were first introduced by Pendry \cite{Pendry-1,Pendry-2},
who suggested that they can also act as perfect lenses \cite{Pendry-3}.

Since the original microwave experiment by Smith {\it et al.}
\cite{Smith-1}, which first materialized Pendry's proposal,
various new samples were prepared \cite{Smith-2,Li} (composed of SRRs
and wires) all of which have been shown to exhibit a pass band in which
it was assumed that  $\eps$ and $\mu$  are both negative.
This assumption was based on measuring independently the transmission, $T$,
of the wires alone, and then the $T$ of the SRRs alone.
If the peak in the combined metamaterial composed of SRRs+wires
were in the stop bands for the wires alone
(which corresponds to negative $\eps$)
{\em and}  for the SRRs alone
(which is thought to correspond to negative $\mu$)
the peak was considered to be left-handed (LH).
Further support to this interpretation was provided by the demonstration
that some
of these materials exhibit negative refraction of electromagnetic
waves \cite{Shelby}.
Subsequent experiments \cite{Claudio-2} have reaffirmed the property of
negative refraction, giving strong support to the interpretation that these
metamaterials can be correctly described by negative permittivity and
negative permeability.
However, as we shall show in the present study, this is not always the case.
The combined system of wires and SRRs exhibits synergy of the two components
as a result of which its effective plasma frequency, $\omega_p'$,
is much lower than the plasma frequency of the wires, $\omega_p$.

There is also a significant amount of numerical work
\cite{MS-2,numerics,numeriks,ms} in which the complex transmission and
reflection amplitudes are calculated for a finite length of metamaterial.
Using these data a retrieval procedure can then be applied
to obtain the effective permittivity $\epsilon$ and permeability $\mu$,
under the assumption that the metamaterial can be treated as
homogeneous.
%
%
This procedure confirmed \cite{SSMS,KMS} that a medium composed of SRRs
and wires could indeed be characterized by effective $\epseff$ and
$\mueff$ whose real parts were both negative over a finite frequency band,
as was the real part of the refractive index $\neff$.

In the present paper, we study the transmission of periodic systems
made up of wires alone, SRRs alone and of combined systems
of wires and SRRs (LHMs).
Through a very detailed retrieval scheme \cite{SSMS}, the effective
$\epseff$ and $\mueff$ of those systems are obtained.
It is shown that the SRRs have also an electric response, in addition to their
magnetic response which was first described by Pendry \cite{Pendry-2}.
The electric response of the SRRs is demonstrated by closing their air gaps,
and therefore destroying their magnetic response.
In fact, it is shown that the electric response of the SRRs is identical to
that of cut-wires.
Analytical expressions for the effective $\eps$ of wires and SRRs
as well as for the effective $\mu$ of SRRs are given.
Using these analytical expressions one is able to reproduce the low frequency
transmission, $T$, and reflection, $R$, characteristics of LHMs.
   Even the minor details in $T$ and $R$ observed in the simulations can be
   analytically explained.
The main power of the present analysis though is that it gives an
easy criterion to identify if an experimental transmission peak is LH or
right-handed (RH):
If the closing of the gaps of the SRRs in a given LHM structure removes
the peak close to the position of the SRR dip from the $T$ spectrum,
this is implies that the $T$ peak is indeed left-handed.
This criterion is very valuable in experimental studies, where one can not
easily obtain the effective $\epseff$ and $\mueff$
and holds provided the magnetic resonance frequency $\omega_m$ is well
separated from $\omega_p'$.
Our criterion is used experimentally and is found that some $T$ peaks that
were thought to be LH, turn out to be right-handed \cite{katsan}.

We use the transfer matrix method to simulate numerically the
transmission properties of the metamaterials.
A representative result is shown in Fig.~\ref{fig1}, presenting
transmission spectra for a usual SRRs+wires metamaterial
and for its isolated constituents.
As expected, the SRRs system (Fig.~1(a)) shows a well defined
$T$ dip near $\omega a/c=0.04$
($a$ is the discretization length, related to the unit cell as explained
in the caption of Fig.~1,
and $c$ is the vacuum speed of light),
due to its magnetic resonance, which gives a negative effective $\mu$
in the dip regime, while its effective $\eps$ is positive  in this
frequency range.
This interpretation has been verified with the retrieval procedure.
The lattice of continuous wires alone shows no transmission  up to the
plasma frequency $\omega_p a/c=0.09$
(see the dotted line in Fig.~1(b))
due to a negative effective $\eps$.
Combining the SRRs and the wires systems we find {what
is shown in Fig.~1(b) - solid line:
no transmission at low frequencies (due to negative $\eps$),
a LH transmission peak (where both $\mu$ and $\eps$ are negative)
that roughly corresponds to the transmission dip of the isolated SRRs,
  followed by
a second transmission gap (where $\mu \ge 0$, $\eps \le 0$)
and terminated by a right-handed
transmission shoulder (at about 0.06). The position
of the last right-handed shoulder, which does not coincide with the
$\omega_p$ of the isolated wires, shows that the SRRs contribute also
to the electric response of the combined
  system, something that has never been considered previously.
This SRR contribution leads to
a new plasma  frequency $\omega_p'$, smaller than $\omega_p$.
The present interpretation of the data is supported again by results
obtained from the retrieval procedure.
Thus, combining the resonant behavior in $\mu(\omega)$ of the SRRs
with the negative behavior of $\eps(\omega)$ due to the wires only
is clearly inadequate; the resonant contribution to  $\eps(\omega)$
due to the SRRs has to be added.

Furthermore, the interaction of the
electric response of wires and SRRs has to be taken also into
account.
Some indication that there exists such a non-negligible electric
interaction between
the SRRs and the wires comes from the comparison of the $T$ for the
in-plane system (i.e. wires next to the SRRs, as in Fig.~1(b)), with the
$T$ for an off-plane system (wires behind SRRs, see Fig.~1(c)-solid line).
It is found that for the in-plane case the  first,
left-handed transmission peak is much narrower and that
the $\omega_p'$ is higher
than for the off-plane case.
The reduction of $\omega_0$ (the electric resonance frequency of the SRRs,
see Eq.~3 below) and consequently of the $\omega_p'$ (Fig.~1(c))
suggests that there is a significant coupling
between the electric response of the wires and that of the SRRs
in the off-plane case.

\begin{figure}
\centerline{\includegraphics[clip,width=7cm]{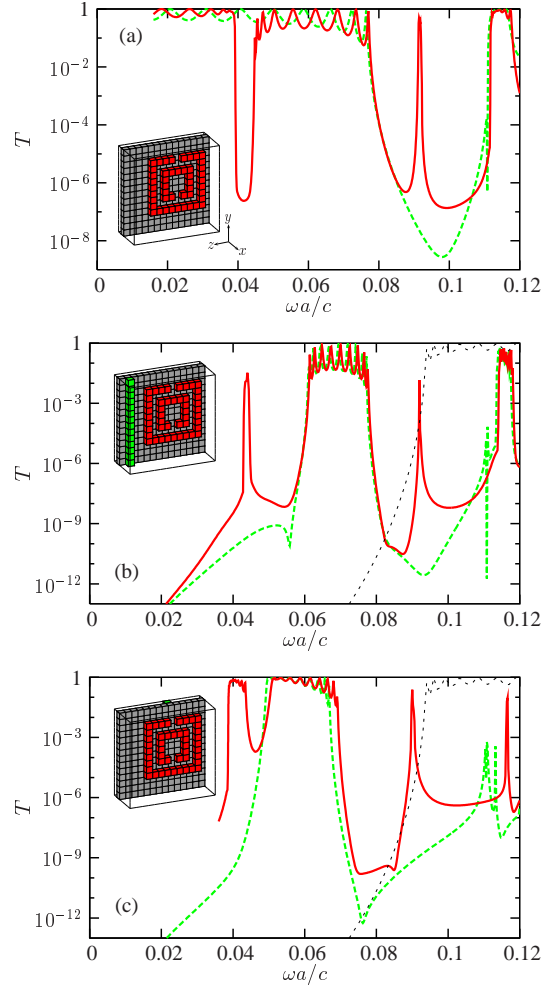}}
\vspace{5mm}
\caption{(a): Frequency dependence of the transmission coefficient of a
lattice of SRRs (solid line) and of closed-SRRs (dashed line);
(b): As in (a) for a metamaterial of SRRs plus wires (solid line) and
of closed-SRRs plus wires (dashed line), with the wires next to the SRRs.
The dotted line shows the transmission for wires only.
(c) As in panel (b) but with the wires in the opposite face
of the SRRs.
The insets show the geometry of the unit cell.
The relative permittivity for the metals is taken to be
$\eps_m = (-3 + i 5.88) 10^5$ and for the dielectric board
$\eps_b = 12.3$.
All relative permeabilities are one.
The unit cell size is 6$a$ x 14$a$ x 14$a$.
$a$ is the discretization length and $c$ the light velocity in air.}
\label{fig1}
\end{figure}

Since all the metals, the vacuum and the dielectric boards are non-magnetic
materials, the only source of major magnetic response of the LHM are the
circulating induced currents in the metallic rings of the SRRs, which are
forced to oscillate due to the gaps acting like capacitors.
Therefore, a simple way to study the combined electric response of the LHM
is to close the  gaps in the rings of the SRRs.
This destroys their magnetic resonance without substantially
affecting their electric response.
Closing the gaps, i.e.  eliminating the capacitors from the rings
of the SRRs, the induced circulating currents inside the SRRs are still
allowed to flow but cannot oscillate independently of the external
electromagnetic (EM) field anymore.
The changes in the electric response from the closing of the SRRs
are expected to be weak (indeed $\omega_0$ moves at most by 3\% as can
be seen in Fig.~1(c)),
since only a very small amount of metal is added (in the gaps of the SRR).
We have checked all these ideas by calculating the transmission
with closed-SRRs.
The results (see the dashed lines in Fig.~1) are almost the same as the
ones  for the open SRRs.
The only important difference is the disappearance of the
first dip in Fig.~1(a), and the first peak in Figs~1(b) and 1(c),
as well as of the peak at about 0.09.
The features that disappeared are due to the magnetic response of the SRRs.
This was also confirmed with our retrieval procedure, which clearly showed
that indeed the dip in Fig.~1(a) and the peaks in Figs 1(b) and 1(c)
are due to the magnetic response of the SRRs.
The most important point of this analysis is that the combination
of the wires with the SRRs can lower, in principle, the $\omega_p$ of
the isolated wires to values very close to the frequency of the
magnetic response, $\omega_m$, of the SRRs.
In some cases it is possible that $\omega_p'$ is lower than the $\omega_m$,
and transmission peaks appear at low frequencies which are not LH but RH.
This has been also confirmed experimentally \cite{katsan}.

Through our detailed transmission studies and the retrieval procedure
for obtaining the effective $\epseff$ and $\mueff$ of the different
structures, we were able to obtain analytic expressions for the effective
$\epseff$ and $\mueff$ for all the structures.
In particular, as expected from effective medium arguments,
an array of metallic wires exhibits the frequency-dependent
plasmonic form
\begin{equation}\label{epseffwire}
\eps^{wire}_{\rm eff}(\omega)=1-\frac{\omega_p^2}{\omega^2+i\omega\gamma}.
\end{equation}

For the frequency dependence of the magnetic response of the SRR we
obtain in the case of a single resonance \cite{SRR} that
\begin{equation}\label{mueff}
\mu^{SRR}_{\rm eff}(\omega)=1-
\frac{\omega_m'^2-\omega_{m}^2}{\omega^2-\omega_{m}^2+i\omega\gamma}.
\end{equation}

We have argued above and we have shown by introducing cut-wires
instead of SRRs that the electric response of the SRRs is equivalent
to that of cut-wires.
Therefore, for the frequency dependence of the electric response of the
SRR we obtain that \cite{SRR}
\begin{equation}\label{epseffSRR}
\eps^{SRR}_{\rm eff}(\omega)=1-
\frac{\omega_p^2-\omega_{0}^2}{\omega^2-\omega_{0}^2+i\omega\gamma}.
\end{equation}

The magnetic resonance of the SRR is due to the oscillation
of circular currents inside the metallic rings,
and is determined by the inductance of the loop (enclosed area)
and the capacitance of the gaps (mainly the gap-width) in the rings.
The second major contribution of the SRR, its electric cut-wire resonance,
is due to the oscillation of  linear currents along
the sides of the SRR which are parallel to the external electric field.
It is basically determined by the size of the SRR
in the direction of the electric field.
In comparison with the response of the continuous wire, Eq.~(1),
for a cut-wire
the electric resonance is shifted from zero to some $\omega_0 \ne 0$
because of the additional depolarization field due to its finite extent.


\begin{figure}
\centerline{\includegraphics[clip,width=7cm]{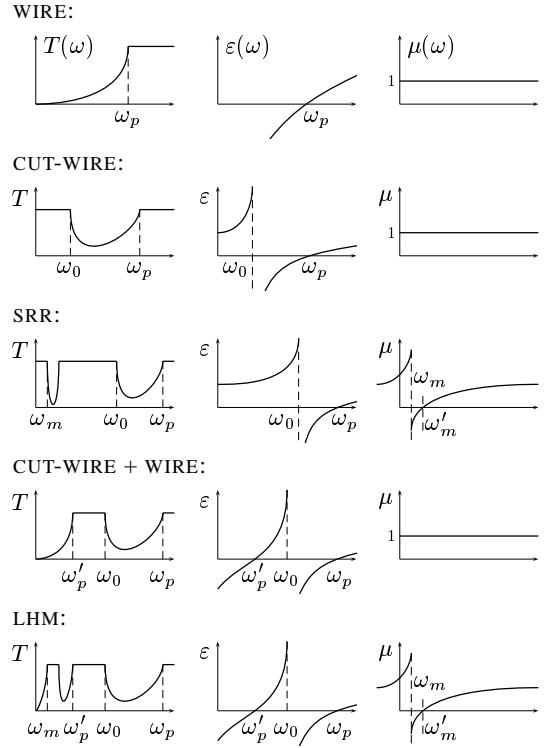}}
\vspace{5mm}
\caption{Each row shows the response of each one of five
periodic systems to electromagnetic wave.
   This response is shown through the frequency dependence of
transmission
coefficient (left panels), dielectric function (middle panels) and
magnetic permeability (right panels).}
\label{fig2}
\end{figure}


The electric response of the LHM is to first approximation equal
to the sum of the electric response of the wires (given by Eq.~(1)) and
the electric cut-wire response of the SRRs (given by Eq.~(3)),
which may be approximated by the closed-SRRs.
The electric interaction between wires and SRRs has to be taken into account
in order to be more realistic.
Since the wires or the cut-wires do not have any magnetic response,
the magnetic response of the LHM is
  just the SRR magnetic resonance, given by Eq.~(2).
In Figure~\ref{fig2} we present in a compact schematic form
the response of the various ``components'' of a metamaterial of
SRRs and wires to an EM field.
This response is shown through the frequency dependence of transmission
coefficient (left panels), dielectric function (middle panels) and
magnetic permeability (right panels).
The first row  shows the response of a periodic system
of infinite wires.
This response is analogous to that of a bulk metal, i.e. there is a cut-off
frequency $\omega_p$ above which $\eps(\omega)$ becomes positive from negative,
and thus the system becomes "transparent" to the EM radiation.
The system does not have any magnetic response.
Second row shows the response of a system of cut-wires
(wires much sorter than the wavelength of the EM wave),
or closed-SRRs.
The difference with the continuous wires is that here the negative $\eps$
regime has also a lower edge $\omega_0\ne 0$,
due to the finite nature of the wires \cite{KMS}.
The $\eps(\omega)$ has the form shown in the middle graph.
Again no magnetic response.
Third row shows the response of a periodic system of
(single-ring) SRRs.
Their electric response is cut-wire like (like in second row) and
their magnetic response has a resonance at $\omega_m$,
where the magnetic permeability ($\mu$) jumps from positive to negative
values.
The transmission (left panel) becomes finite in the regions of
positive product $\eps \mu$ and goes to zero for negative $\eps \mu$.
The relative order of the characteristic frequencies
($\omega_m$, $\omega_p$, $\omega_0$) depends on the parameters of
each specific system.
Fourth row shows the response of a system of infinite wires
plus cut-wires (closed-SRR).
This system contains in fact all the electric response of the LHM,
without its magnetic response.
The combined electric response of wires and cut-wires
leads to a new cut-off frequency, $\omega_p'$,
much lower than $\omega_p$.
Last row shows the full response of the LHM; its electric response is
that of a system of wires plus cut-wires (fourth row) and its magnetic
response is that of a periodic system of SRRs (third row).
Simulations (see Fig.~1) and the retrieval procedure have verified all
aspects shown schematically in Fig.~2
(for the explanation of the peak at 0.09 see Ref.~\cite{SRR}).

We have systematically studied the transmission properties of LHMs
composed of SRRs and continuous wires,
and  also of their various components separately.
A retrieval procedure was used to obtain the effective parameters
for each case.
We found that the electric response of a LHM is the sum of electric
responses of the wires and the SRRs.
This changes the existing picture where the electric response of a LHM
was attributed only to the wires, and provides a valid scheme for
interpreting  experimentally observed transmission peaks.

\medskip

\noindent Acknowledgments. This work was partially supported by Ames Laboratory
(Contract. n. W-7405-Eng-82). Financial support of
EU$\underline{~~}$FET project DALHM, NSF (US-Greece Collaboration),
and DARPA (Contract n.
MDA972-01-2-0016)
   are also acknowledged.



\begin{thebibliography}{99}
\bibitem{SRR} The conventional SRR shown in Fig.~1(a) has two
rings and therefore two magnetic resonances.
The peak shown in Figs 1 (solid lines) around 0.09 is
due to the magnetic resonance of the inner ring.
For the same reasons we expect a second electric cut-wire resonance
at higher frequencies.
In addition to the capacitance provided by the gaps in the rings
there is a inter-ring capacitance, which modifies the $\omega_m$ of the
isolated rings.
\end{thebibliography}
\end{document}